\documentclass[12pt]{article}

\newcommand{\<}{\langle}

\newcommand{\F}{{\Phi}}

\newcommand{\G}{\Gamma}
\newcommand{\HH}{{\cal H}}

\newcommand{\rhsp}{\F_+\subset\HH\subset\F^\times_+}
\newcommand{\rhsm}{\F_-\subset\HH\subset\F^\times_-}

\newcommand{\om}{\omega}

\newcommand{\C}{{\rm {l\!\!\! {C}}}}
\newcommand{\CS}{{\cal S}}

\newcommand{\Bigtriangleright}{{\displaystyle{{\rhd}}}}
\newcommand{\Bigtriangleleft}{{\displaystyle{{\lhd}}}}

\newcommand{\hin}{\rangle\!}
\newcommand{\vor}{\!\langle}

\usepackage{latexsym}
\usepackage{amsmath, amssymb}
\usepackage[]{graphicx}
\usepackage{epsf}
\usepackage{subeqnarray}
\begin{document}

\def\theequation{\thesection.\arabic{equation}}
\setcounter{equation}{0}

\title{ \vspace{-.5in}{\bf Higher Order Gamow States
with Exponential Decay}} \vspace{4pt}
\author{
		{\bf Christoph P{\"u}ntmann}  \vspace{4pt}\\ 
		\footnotesize{The University of Texas at Austin}\\
		\footnotesize{Austin, Texas~78712-1081}
}
\maketitle

\begin{abstract}
We derive Gamow vectors from S-matrix poles of higher multiplicity in
analogy to the Gamow vectors describing resonances from first-order
poles. With these vectors we construct a density operator that 
describes resonances associated with higher order poles that
obey an exponential decay law. It turns out that this
operator formed by these higher order Gamow vectors has a unique structure.
\end{abstract}

\newpage
\section{Introduction}\label{sec:introduction}
\setcounter{equation}{0}

Resonances in quantum mechanics can be described by poles of the
analytically continued S-matrix on the second sheet of a two-sheeted
Riemann surface (Newton, 1982; Goldberger and Watson, 1964; 
Bohm 1979, 1980, 1981, 1993).
These poles appear in conjugate pairs below (at
$z_R=E_R-i\Gamma/2$ ) and above (at $z_R^*=E_R+i\Gamma/2$) the
positive real axis, where the pole at $z_R$ corresponds to a decaying
state at times $t\geq0$ and the pole at $z_R^*$ to the growing state
at $t<0$. These poles lead to the description of the Gamow
vectors with energy $E_R$ and lifetime $\tau=\hbar/\Gamma$. 
The Gamow vectors possess all the properties of resonances, in particular,
an exponential decay law and a Breit-Wigner energy distribution.
One can extend this derivation of the Gamow vectors from first order
poles to S-matrix poles of higher
multiplicity (Antoniou and Gadella, 1995; Bohm {\em et al.}, 1997). 
An S-matrix pole of
order $r$ at the position $z_R=E_R-i\Gamma/2$ on the second Riemann
sheet leads to a set of $r$ generalized eigenvectors of the
Hamiltonian of order
$k=0,~1,\dots,~r-1$, which are Jordan vectors of degree $k+1$ to the
generalized eigenvalue $E_R-i\Gamma/2$ and which are elements of a
generalized complex eigenvector expansion (nuclear spectral theorem
in the rigged Hilbert space).
The form of this generalized complex eigenvector expansion
suggests the definition of a state operator (density matrix) for the
microphysical decaying 
state from a higher order pole. This microphysical state is not a pure
state, but a mixture of
non-reducible components. In spite of the fact that the $k$-th order
Gamow-Jordan vectors have a polynomial time-dependence besides the
exponential, which in the past were 
always associated with resonances from higher order poles, 
this microphysical state obeys a purely exponential decay law.

Resonances from higher order poles, in particular double poles, have
already been described 
about 30 years ago (Goldberger and Watson; Newton, 1982; Goldhaber, 1968), 
but were always associated with an additional polynomial
time dependence not confirmed in
experiment. However, operators containing
finite dimensional matrices consisting of non-diagonalizable Jordan
blocks have been discussed in connection with resonances numerous
times in the
past (Mondragon, 1994; Stodolsky, 1970; Lukierski, 1967; Dothan and
Horn, 1970;
Katznelson, 1980; Bhamathi and Sudarshan, 1995; Br\"{a}ndas and
Dreismann, 1987; Antoniou and Tasaki, 1993).
Using the formalism of the rigged
Hilbert space, I.~Antoniou and M.~Gadella derived the Gamow-Jordan
vectors, or higher order Gamow vectors, from the higher order
poles of the S-matrix.

In the second section of this article, we recall some of the
notation (Bohm, 1993; Bohm {\em et al.}, 1997) needed for the description of 
the scattering experiment in the rigged
Hilbert space formalism. We show how to obtain these
hypothetical vectors associated with the higher order S-matrix
poles and show that they are Jordan vectors (Baumg\"{a}rtel, 1984; 
Kato, 1966; Gantmacher, 1959). 
We derive their properties under the action of the
Hamiltonian and their semigroup time evolution.
In the third section, we discuss possible operators formed by these
vectors to represent microphysical states describing resonances.
In the fourth section, we will ask for the converse : Going out
from an exponential decay law for the time evolution of 
resonances from higher order
S-matrix poles, what is the most general form of the operator 
formed by dyadic products of higher order Gamow vectors?

\section{Poles of the S-Matrix and Gamow-Jordan Vectors}
\label{sec:poles}
\setcounter{equation}{0}

We recall that in the rigged Hilbert space formalism one
uses two different space triplets for the set of in-states 
$\phi^+\in\; \rhsm$ describing the preparation process of the
scattering experiment and the set of observables $\psi^-\in\;\rhsp$
describing the registration process (Bohm {\em et al.}, 1997; Bohm and
Gadella, 1989)
The in-state $\phi^+$, which evolves from the
prepared in-state $\phi^{\text{in}}$ outside the interaction region,
is determined by the accelerator. The so-called
out-state $\psi^-$ (or $\psi^{\text{out}}$) is determined by the 
detector. $|\psi^{\text{out}} \rangle \langle\psi^{\text{out}} |$ 
is therefore the observable which the detector registers and not 
a state. The S-matrix elements are given by the projection of the set
of out-states $\{\phi^{\rm out}\}$ onto the set of observables 
$\{\psi^{\rm out}\}$ (Bohm, 1993)
\begin{equation}
	\left(\psi^{\text{out}},\phi^{\text{out}}\right)=
        \left( \psi^{\text{out}},S \phi^{\text{in}}\right) 
	= \left( \psi^-,\phi^+\right) 			\label{eq:outmatrix}
  	=\int_{\text{spec}\,H}dE\;\langle\psi^-|E^-\rangle 
        S(E)\langle^+  E|\phi^+\rangle\;.
\end{equation}
The vectors $|E^\pm\rangle\in\Phi^\times$ are the scattering states
(Dirac kets) and are eigenvectors of the exact Hamiltonian with energy
label $E$, which can take values on a two-sheeted Riemann surface.
We choose to ignore all other labels of the basis vectors
$|E^\pm\rangle$, since nothing important is gained in our discussion
if we retain the additional quantum numbers, e.g., the angular momentum
quantum numbers $l$ and $l_3$ or the polarization or channel quantum
numbers $\eta$.
Thus, we shall restrict our discussion on one initial channel 
$\eta =\eta_A$ and one final channel $\eta'=\eta_B$ 
(e.g., $\eta_B=\eta_A$ for elastic scattering), and we shall
consider the $l$-th partial wave of the $\eta_B$-th channel (Bohm, 1993):
i.e., $S(E)\equiv S^{\eta_B}_l (E)$.
We consider the model in which the S-matrix is analytically
continued to a two-sheeted Riemann surface in the energy
representation (Bohm, 1993) and in which the S-matrix $S(\omega)$,
$\omega\in\C$,
has one $r$-th order pole at the position $\omega=z_R$ ($z_R =E_R-i\,
\G /2$) in the lower half-plane of the second sheet, (and consequently
there is also one $r$-th order pole in the upper half-plane of the
second sheet at $\omega=z^\ast_R$). In this paper we will not discuss
the pole at $z^\ast_R$, as it leads to $r$ growing higher order Gamow vectors, 
and the correspondence between the growing and decaying vectors is
just the same as for first-order pole resonances ($r=1$). 
The model that we discuss here 
can easily be extended to any finite number of finite order poles in
the second sheet below the positive real axis.\\

The unitary S-matrix of a quasistationary state associated with an
$r$-th order pole at $z_R=E_R-i\Gamma/2$ in the
lower half-plane of the second sheet (denoted by II) is represented 
by (Bohm, 1993, sect.~XVIII.6)
\begin{equation}
	S_{\rm II}(\omega)=e^{2i\delta_R(\omega)}e^{2i\gamma(\omega)}\;,
\end{equation}
where $\delta_R(\omega)=r\,{\rm arctan}(\frac{\Gamma}{2(E_R-\omega)})$
is the rapidly varying resonant part of the phase shift, and
$\gamma(\omega)$ is the background phase 
shift, which is a slowly varying function of the complex energy
$\omega$. $r$ is a dimensionless quantity that, due to the analyticity
properties of the S-matrix (Bohm, 1993, section XVIII.6), 
takes integer values, where $r>0$ leads to a decaying resonance of order $r$, 
and $r<0$ to its corresponding growing state. Using the identity $\;{\rm
arctan}\frac{\Gamma/2}{E_R-\omega}=\frac{i}{2}
\left({\rm ln}(\omega-E_R-i\Gamma/2)-
{\rm ln}(\omega-E_R+i\Gamma/2)\right)$, one rewrite $S_{\rm
II}(\omega)$:
\begin{eqnarray}\label{b16a}\label{b16c}
	S_{\rm II}(\omega)
	&=&\left(\frac{\omega-E_R-i\Gamma/2}{\omega-(E_R-i\Gamma/2)}
	\right)^re^{2i\gamma(\omega)}
	=e^{2i\gamma(\omega)}+\sum_{l=1}^{r}\begin{pmatrix}r\\l\end{pmatrix}
	\frac{(-i\Gamma)^l}{(\omega-z_R)^l}e^{2i\gamma(\omega)}
\end{eqnarray}
We insert this into~(\ref{eq:outmatrix}) and deform the contour of
integration ${\cal C}_-$ through the cut along the spectrum of $H$
into the second sheet (Bohm, 1979, 1980, 1981, 1993). Then one obtains
\begin{eqnarray}\label{paul17}\label{bohm17}
	(\psi^-,\phi^+)&=&\int_{{\cal C}_-}\,d\om \;\langle\psi^-|\om^-
	\rangle\,S_{\rm II}(\omega)\,\langle^+\om |\phi^+\rangle+\\\nonumber
	&&+\sum_{n=0}^{r-1}\begin{pmatrix}r\\n+1\end{pmatrix}
	(-i\Gamma)^{n+1}\oint_{\hookleftarrow} d\om\, 
        \langle\psi^-|\om^-\rangle\,\frac{e^{2i\gamma(\omega)}}
	{(\om-z_R)^{n+1}}\,\langle^+\om|\phi^+\rangle;
	\hspace{.2in}{\rm Im}(\omega)<0\;. 
\end{eqnarray}
The first integral does not depend on the pole and is called the
``background term''. The contour ${\cal C}_-$ can be deformed into the
negative axis of the second sheet from 0 to $-\infty_{\rm II}$. 
\begin{equation}\label{eq:regular}
	(\psi^-,\phi^+)=
	\int_0^{-\infty_{\rm II}}dE\langle\psi^-|E^-\rangle\,S_{\rm
	II}(E)\,\langle^+E|\phi^+\rangle\,+\,(\psi^-,\phi^+)_{\rm P.T.}
\end{equation}
We will not need to further investigate the background integral in this
presentation. For the higher
order pole term $(\psi^-,\phi^+)_{\rm P.T.}$ we obtain using the
Cauchy integral formulas $\oint_{\hookrightarrow}{\frac{f(\om )}{(\om
-z_R)^{n+1}}}\;d\om={\frac{2\pi i}{n!}}\left. f^{(n)}(z)\right|_{z=z_R}$ 
where $f^{(n)}(z)\equiv{\frac{d^nf(z)}{dz^n}}$:
\begin{eqnarray}\label{bohm18}
	(\psi^-,\phi^+)_{\rm P.T.}
	&=& \sum_{n=0}^{r-1}\begin{pmatrix}r\\n+1\end{pmatrix}
	(-i\Gamma)^{n+1}\left(-\frac{2\pi i}{n!}\right)
   	\Bigl(\langle \psi^-|\om^-\rangle\; e^{2i\gamma(\omega)}
	\;\langle^+ \om |\phi^+\rangle\Bigr)^{(n)}_{\om =z_R}
\end{eqnarray}
where $\left(\dots\right)^{(n)}_{\omega=z_R}$ denotes the $n$-th
derivative with respect to $\omega$ taken at the value $\omega=z_R$.

Since the kets $|\omega^-\rangle$ are (like the Dirac kets
$|E^-\rangle$) only defined up to an arbitrary factor or, if their
``normalization'' is already fixed, up to a phase factor, we can absorb the
background phase $e^{2i\gamma(\omega)}$ into the kets
$|\omega^-\rangle$ and define
\begin{equation}\label{b18a}
	|\omega^\gamma\rangle\equiv|\omega^-\rangle 
	e^{2i\gamma(\omega)}\;.
\end{equation}
Note that this phase is not trivial, e.g. $|E^+\rangle=|E^-\rangle
S_{\rm II}(E)=|E^-\rangle e^{2i\delta_R(E)}e^{2i\gamma(E)}$, 
except for the case when the slowly varying
background phase $\gamma(\omega)$ is constant and the
$|\omega^\gamma\rangle$ are identical with $|\omega^-\rangle$ 
up to a totally trivial constant phase factor. In
general~(\ref{b18a}) is a non-trivial gauge transformation. 
We will keep the phase $\gamma$ in our Gamow vectors, but not
investigate their properties further. The results of this section are
true also for the case when $|z_R^\gamma\hin^{(k)}$ are exchanged by
$|z_R^-\hin^{(k)}$. But as we cannot just ignore the existence of the
background integral in~(\ref{bohm17}), one has to keep in mind that 
their existence is not irrelevant, if one deals with poles of order $r>1$.

Taking the derivatives, we write the pole term as
\begin{equation}\label{bohm41}
	(\psi^-,\phi^+)_{\rm P.T.}=\sum_{n=0}^{r-1}
	\begin{pmatrix}r\\n+1\end{pmatrix}(-i\Gamma)^{n+1}\left(
	-\frac{2\pi i}{n!}\right)\sum_{k=0}^n
	\begin{pmatrix}n\\k\end{pmatrix}\langle \psi^-|
	z_R^\gamma\hin^{(k)}\,^{(n-k)}\vor^+z_R|\phi^+\hin\;,
\end{equation}
where we denote the 
$n$-th derivative of the analytic function
$\langle\psi^-|z^\gamma\rangle$ by
$\langle\psi^-|z^\gamma\hin^{(n)}$ with value
$\langle\psi^-|z^\gamma_R\hin^{(n)}$ at $z=z_R$.
Since $\langle\psi^- |E^-\rangle \in \CS\cap\HH^2_-$, i.e. element of
the Schwartz space and of Hardy class (Duren, 1970; Hoffman, 1962;
Bohm and Gadella, 1989), it follows that $\langle\psi^-
|z^\gamma\hin^{(n)}$ is also an analytic function in the
lower half-plane of the second sheet, whose boundary value on the positive
real axis $\langle\psi^-|E^\gamma\hin^{(n)}\in \CS\cap\HH^2_-$.
Analogously,  we denote by $\,^{(n)}\vor^+ z|\phi^+\rangle$ the 
$n$-th derivative of the analytic function $\langle^+ z|\phi^+\rangle$.
Again, $\,^{(n)}\vor^+ z|\phi^+\rangle$ is analytic in the lower
half-plane with its boundary value on the real axis being 
$\,^{(n)}\vor^+E|\phi^+\rangle\in\CS \cap\HH^2_-$.
\\
The $r$-th order pole is therefore associated 
with the set of $r$ generalized vectors 
\begin{equation}\label{bohm38}
	|z^\gamma_R\hin^{(0)}\,,\;|z^\gamma_R\rangle^{(1)}\,,\;\cdots,
	|z^\gamma_R\hin^{(k)}\,,\cdots,|z^\gamma_R\hin^{(n)}\;.
\end{equation}
For the first order resonance pole this, of course, reduces to the
single vector $|z_R^\gamma\rangle=|z_R^\gamma\hin^{(0)}$ in agreement
with (Bohm, 1979, 1980, 1981, 1993).

We can now establish the complex basis vector expansion in analogy to
the Dirac basis vector expansion (Nuclear Spectral Theorem in the
rigged Hilbert space, (Gel'fand and Vilenkin, 1964; Bohm and Gadella, 1989)). 
If we return to the complete S-matrix
element~(\ref{bohm17}) and insert the pole term, we get 
\begin{eqnarray}\label{bohm43}
	(\psi^-,\phi^+)&=&\int_0^{-\infty_{\rm
	II}}dE\,\langle\psi^-|E^+\rangle\langle^+E|\phi^+\rangle\\\nonumber
	&&\!\!\!-\sum_{n=0}^{r-1}
	\begin{pmatrix}r\\\!n\!+\!1\!\end{pmatrix}
	\frac{2\pi\Gamma}{n!}(-i\Gamma)^n
	\sum_{k=0}^n\begin{pmatrix}n\\k\end{pmatrix}
	\langle\psi^-|z^\gamma_R\hin^{(k)}
	\;^{(n-k)}\vor^+\!z_R|\phi^+\rangle			
\end{eqnarray} 
Omitting the arbitrary $\psi^-\in\Phi_+$ and
rearranging the sums in the second term, we obtain the complex basis 
vector expansion for an arbitrary $\phi^+\in\Phi_-$
\begin{equation}\label{bohm44}
	\phi^+
	=\int_0^{-\infty_{\rm II}}dE|E^+\rangle\langle^+E|
	\phi^+\rangle\,+\,\sum_{k=0}^{r-1}\,b_k
	|z_R^\gamma\hin^{(k)}\;,
\end{equation}
where the coefficients $b_k$ are given by
\begin{equation}
	b_k=(-2\pi\Gamma)\sum_{n=k}^{r-1}
	\begin{pmatrix}r\\n+1\end{pmatrix}
	\begin{pmatrix}n\\k\end{pmatrix}\frac{(-i\Gamma)^n}{n!}
	\,^{(n-k)}\vor^+z_R|\phi^+\rangle
\end{equation}
This complex
generalized basis vector expansion is the most important result of our
irreversible quantum theory (as is the Dirac basis vector expansion
for reversible quantum mechanics).
\noindent It shows that the generalized vectors~(\ref{bohm38})
(functionals over the space $\Phi_+$) are part of a basis system for
the $\phi^+\in\Phi_-$ and form together with the kets
$|E^+\rangle,~~-\infty_{\rm II}<E\leq0$, a complete basis system. The 
vectors~(\ref{bohm38}) span a linear subspace ${\cal
M}_{z_R}\subset\Phi^\times_+$ of dimension $r$:
\begin{equation}\label{bohm48}
	{\cal M}_{z_R} = \Biggl\{ \; \xi \; {\Bigg|} \; 
  	\xi =\sum^{r-1}_{k=0}\zeta_k|z^\gamma_R \hin^{(k)}\,;\; \zeta_k\in\C\, 
	\Biggr\} \subset\F^\times_+ 
\end{equation}
If there are $N$ poles at $z_{R_i}$ of order $r_i$, then for every
pole there exists a linear subspace ${\cal M}_{z_{R_i}}\subset\Phi^\times_+$.

Note that the label $k$ of the higher order Gamow vectors is not a quantum
number in the usual sense. Basis vectors are usually labeled by
quantum numbers associated with eigenvalues of a complete system of
commuting observables (Bohm, 1993, chap.~IV). But there is no
physical observable that the label $k$ is connected to. Therefore, the
different $|z^\gamma_R\hin^{(k)}$ in the subspace ${\cal M}_{z_R}$
do not have a separate physical meaning.

Now that~(\ref{bohm44}) has established the generalized vectors~(\ref{bohm38})
as members of a basis system (together with the
$|E^+\rangle;~~-\infty_{\rm II}<E\leq0$) in $\Phi^\times_+$, we can obtain
the action of the Hamiltonian $H$ by the action of the operator
$H^\times$ on these basis vectors. We will write this Hamiltonian in
terms of its matrix elements in this basis. 
For this purpose we replace the arbitrary $\psi^-\in\Phi_+$
in~(\ref{bohm43}) by $\tilde{\psi}^-=H\psi^-$ which is again an
element of $\Phi_+$, and one finds (Antoniou and Gadella, 1995; Bohm
{\em et al.}, 1997)
\begin{eqnarray}\nonumber
	\langle\tilde{\psi}^-|z^\gamma_R\hin^{(k)}&\equiv&
	\langle H\psi^-|z^\gamma_R\hin^{(k)}\equiv\langle\psi^-|H^\times 
	|z^\gamma_R\hin^{(k)}\\				
	&=&z_R\langle\psi^-|z^\gamma_R\hin^{(k)}+\langle\psi^-
	|z^\gamma_R\hin^{(k-1)}\;;\hspace{.3in}k=1,\dots,r-1\;,\nonumber\\
	&&\label{b56}\\\nonumber
	\langle\tilde{\psi}^-|z^\gamma_R\hin^{(0)}&\equiv&
	\langle H\psi^-|z^\gamma_R\hin^{(0)}=\langle\psi^-|H^\times 
	|z^\gamma_R\hin^{(0)}=z_R\langle\psi^-|z^\gamma_R\hin^{(0)}\;.
\end{eqnarray}
This can also be written as a set of functional equations over $\F_+$:
\begin{eqnarray}
	H^\times |z^\gamma_R\hin^{(k)}&=&z_R|z^\gamma_R\hin^{(k)}
	+k|z^\gamma_R\hin^{(k-1)}\,,\hspace{.3in}k=1,\dots,r-1,\nonumber\\
	H^\times |z^\gamma_R\rangle\hspace{4mm}&=&z_R|z^\gamma_R\rangle\,.	\label{b57}
\end{eqnarray}
This means that $H^\times$ restricted to the subspace ${\cal M}_{z_R}$
is a Jordan operator of degree $r$, and the vectors
$|z^\gamma_R\hin^{(k)},~k=0,~1,~2,\dots,~r-1$ are Jordan
vectors of degree $k+1$ (Baumg\"{a}rtel, 1984; 
Kato, 1966; Gantmacher, 1959).
They fulfill the generalized eigenvector equation (Lancaster and
Tismenetsky, 1985),
\begin{equation}
	(H^\times -z_R)^{k+1}|z_R^\gamma\hin^{(k)} =0\;.
\end{equation}
Since, according to~(\ref{bohm44}), the basis system also includes 
the $|E^+\rangle,~-\infty_{\rm II}<E\leq0$, we indicate this by a
continuously infinite diagonal matrix equation
\begin{equation}\label{b59}
	\Bigl(\langle H\psi^-|E^+\rangle\Bigr)=
	\Bigl(\langle\psi^-|H|E^+\rangle\Bigr)=
	\Bigl(E\Bigr)\Bigl(\langle\psi^-|E^+\rangle\Bigr) 
\end{equation}
where $\Bigl(\langle\psi^-|E^+\rangle\Bigr)$ indicates a continuously
infinite column matrix. Then~(\ref{b56}) can be rewritten as:
\begin{eqnarray}
      \left( \begin{matrix}\<\psi^-|\,H^\times\,|z^\gamma_R\hin^{(0)} \\
                \<\psi^-|\, H^\times \, |z^\gamma_R\hin^{(1)} \\
                   \;\vdots\\
		   \;\vdots\\
                \<\psi^-|\, H^\times \, |z^\gamma_R\hin^{(r-1)} \\
         \<\psi^-|\, H^\times \, |E^+ \rangle
	\end{matrix} \right)
      \label{eq:54}\label{paul49}\label{b60}
      &=&\begin{pmatrix} z_R &0&0&\ldots&0&0\\
                1&z_R &0&\ldots&0&\\
                0&2&z_R&\ldots&0&\vdots\\       
                \vdots&\vdots&\ddots &\ddots&\vdots&\\
                0&0&\ldots&r\!-\!1&z_R &0\\
          0&&\ldots&&0&\left( E\right) 
	\end{pmatrix}
 	\begin{pmatrix} \<\psi^-|z^\gamma_R\hin^{(0)}\\
                 \<\psi^-|z^\gamma_R\hin^{(1)}\\
                    \;\vdots\\
                    \;\vdots\\
                 \<\psi^- |z^\gamma_R\hin^{(r-1)}\\
                 \<\psi^- |E^+\rangle 
	\end{pmatrix}\hspace{.5in}
\end{eqnarray}
In this matrix representation of $H^\times$, the upper left
$r\times r$ submatrix associated with the complex
eigenvalue $z_R$ is a (lower) Jordan block of degree $r$.
One can attain the standard form of a Jordan block with 1's on the
lower diagonal by simply choosing the normalization 
\begin{equation}\label{bohm40}
	|z^\gamma_R\hin^{(k)}\rightarrow\frac{1}{k!}|z^\gamma_R\hin^{(k)}
	\hspace{.3in}{\rm and}\hspace{.3in}
	\,^{(l)}\vor^+z_R|\rightarrow\;^{(l)}\vor^+z_R|\frac{1}{l!}\;.
\end{equation}

Next, we are going to discuss the time evolution of the higher order
Gamow vectors.
We replace the arbitrary $\psi^-\in\Phi_+$
in~(\ref{bohm43}) by $\tilde{\psi}^-=e^{iHt}\psi^-$.
We recall that $e^{iHt}$ 
needs to be a continuous operator with respect to the topology
$\tau_{\F_+}$ of the space $\Phi_+$, and its values
$e^{izt}$ need to be holomorphic in all $\Phi_+$. 
Its conjugate $(e^{iHt})^\times$ which acts on the vectors
$|\omega^-\rangle\in\Phi_+$ is only defined 
for positive values of the parameter
$t$ (semigroup time evolution),
\begin{equation}
	\langle e^{iHt}\psi^-|\omega^-\rangle=
	\langle\psi^-|\left(e^{iHt}\right)^\times|\omega^-\rangle
	=e^{-i\omega t}\langle\psi^-|\omega^-\rangle\;,
	\hspace{.2in}t\geq0\;,
\end{equation}
for all $|\omega^-\rangle\in\Phi_+^\times$. 
Then
\begin{eqnarray}\nonumber
	\langle e^{iHt}\psi^-|z^\gamma_R\hin^{(k)}
	&=&\frac{d^k}{d\omega^k}\Big(\langle e^{iHt}\psi^-|\omega^-\rangle
	e^{2i\gamma(\omega)}\Big)_{\omega=z_R}
	=\frac{d^k}{d\omega^k}\Big(e^{-i\omega t}\langle\psi^-|
	\omega^-\rangle e^{2i\gamma(\omega)}\Big)_{\omega=z_R}
	\\&=&
	e^{-iz_Rt}\sum_{p=0}^k
	\begin{pmatrix}k\\p\end{pmatrix}(-it)^p
	\langle\psi^-|z^\gamma_R\hin^{(k-p)}\;.
\end{eqnarray}
This can be written as a functional equation as
\begin{equation}\label{time1}
	\left( e^{iHt}\right)^\times|z^\gamma_R\hin^{(k)}=
	e^{-iz_Rt}\sum_{p=0}^k\begin{pmatrix}k\\p\end{pmatrix}(-it)^{k-p}\;
	|z^\gamma_R\hin^{(p)}\;.
\end{equation}
In the same way one derives for the complex conjugate:
\begin{equation}\label{time2}
	\,^{(l)}\vor^\gamma z_R|e^{iHt}=e^{iz_R^*t}\sum_{q=0}^l
	\begin{pmatrix}l\\q\end{pmatrix}(it)^{l-q}\;^{(q)}\vor^\gamma z_R|\;.
\end{equation}
The same formulas apply also to the vectors
$|z^-_R\hin^{(k)}$ and $^{(l)}\vor^-z_R|$ with background phase $\gamma=0$.
It is important to note that the time evolution operator 
$\left(e^{iHt}\right)^\times $
transforms between different $|z^\gamma_R\hin^{(k)}$, or different 
$|z^-_R\hin^{(k)}$, $k=0 ,\, 1 \ldots, n$, 
that belong to the same pole of order $r$ at $z=z_R$,
but the time evolution does not transform out of ${\cal M}_{z_R}$.

\section{States from Higher Order Gamow Vectors}
\label{sec:states}
\setcounter{equation}{0}

Gamow states of zeroth order with their empirically well-established
properties (exponential time evolution, Breit-Wigner energy
distribution) have been abundantly observed in nature as resonances
and decaying states. Theoretically, there should be no reason why
{quasi}{stationary} states (i.e. states that also cause large time delay
in a scattering process (Bohm, 1993, chap.~18) associated
with integers $r>1$ in~(\ref{b16a}) should not exist. However, no such
quasistationary states have so far been established empirically. One
argument against their existence was that the polynomial time dependence,
which was always vaguely associated with higher order
poles (Goldberger and Watson; Newton, 1982; Goldhaber, 1968), 
and which has not been observed for quasistationary states.
The question that we want to discuss in this section is, whether there is
an analogous physical interpretation for the higher order Gamow vectors
as for the ordinary Gamow vectors, namely as states
which decay (for $t>0$) or grow (for $t<0$) in one prefered direction
of time (``arrow of time'') and obey the exponential law.

In analogy to von Neumann's definition of a pure stationary state
using dyadic products $|f\rangle\langle f|$ of the energy eigenvectors
$|f\rangle$ in Hilbert space, microphysical Gamow states connected
with first order poles can be defined as dyadic products of zeroth
order Gamow vectors (Bohm, 1979, 1980, 1981, 1993; Bohm {\em et al.},
1997),
\begin{equation}\label{g60}
	W^{(0)}=|z^-_R\rangle\langle^- z_R|\;.
\end{equation}
(Note that this definition is not related to the scattering background
phase $\gamma$ which entered in~(\ref{b18a}).)
The time evolution of this Gamow state is exponential, 
\begin{eqnarray}
	W^G(t)&\equiv&\left(e^{iHt}\right)^\times\,|z^-_R\rangle
	\langle^- z_R|\,e^{iHt} \nonumber\\ 	\label{WG62}\label{g61}
	&=&e^{-iz_R t}\,|z^-_R\rangle\langle^- z_R|\,e^{iz^*_Rt}\\
 	&=&e^{-i(E_R-i(\Gamma/2))t}|z^-_R\rangle\langle^- z_R|
	\,e^{i(E_R+i(\Gamma/2))t}\nonumber\\
	&=&e^{-\Gamma t}\,W^G(0)\;,\hspace{4.5cm}t\geq0\;.\nonumber 
\end{eqnarray}
Mathematically, equation~(\ref{g61}) is understood as the
functional equation of
\begin{eqnarray}\label{g62}\label{WG63}
	\langle\psi^-|W^G(t)|\psi^-\rangle
	&=&e^{-\Gamma t}\langle\psi^-|W^G|\psi^-\rangle\hspace{0.2in}
	{\rm for~all}\quad\psi^-\in\Phi_+
	\quad{\rm and}\quad t\geq0.\nonumber
\end{eqnarray}
This shows how important it is in the RHS formalism of quantum
mechanics to know what question one wants to ask about a microphysical 
state when one makes the hypothesis~(\ref{g60}). The vectors
$\psi^-\in\Phi_+$ represent observables defined by the detector 
(registration apparatus), and therefore the
operator $W^G$ represents the microsystem that affects the
detector. Therefore the quantity $\langle\psi^-|W^G|\psi^-\rangle$ is
the answer to the question:
{\it What is the probability that the
microsystem affects the detector?}\\
If the detector is triggered at a later time
$t$, i.e. when the observable has been time translated 
\begin{equation}\label{g63}
	|\psi^-\rangle\langle\psi^-|\quad\longrightarrow\quad
	e^{iHt}|\psi^-\rangle\langle\psi^-|e^{-iHt}\,
	=\,|\psi^-(t)\rangle\langle\psi^-(t)|
\end{equation}
then the same question for $t\geq0$ has the answer: The probability
that the microsystem affects the detector at $t>0$ is
\begin{eqnarray}\nonumber
	\hspace{-.3in}\langle\psi^-(t)|W^G|\psi^-(t)\rangle
	&=&\langle e^{iHt}\psi^-|W^G|e^{iHt}\psi^-\rangle\\\label{g64}
	&=&\langle\psi^-|\left(e^{iHt}\right)^\times 
	W^Ge^{iHt}|\psi^-\rangle\\\nonumber
	&=&e^{-\Gamma t} \langle \psi^-| W^G |\psi^-\rangle\;.
\end{eqnarray}
This means that~(\ref{g64}) is the probability to observe the decaying
microstate at a time $t$ relative to the probability
$\langle\psi^-|W^G|\psi^-\rangle$ at $t=0$ (which one can
``normalize'' to unity by choosing the appropriate factor on the
right-hand side of~(\ref{g60})).

The question that one asks in the scattering experiment is
different. There the pole term (P.T.) of~(\ref{eq:regular}) 
for $r=1$ describes how the microsystem propagates the effect which the
preparation apparatus (accelerator, described by the state $\phi^+$)
causes on the registration apparatus (detector, described by the
observable $\psi^-$).
This involves both the observables $\psi^-\in\Phi_+$ and the prepared
states $\phi^+\in\Phi_-$, and one would ask the question: {\it What is the
probability to observe $\psi^-(t)$ in a microphysical resonance state
of a scattering experiment with the prepared in-state $\phi^+$?}\\
In distinction to the decay experiment where one just asks
for the probability of $\psi^-\in\Phi_+$, in the resonance scattering
experiment one asks for the probability that relates
$\psi^-\in\Phi_+$ to $\phi^+\in\Phi_-$ via the microphysical resonance
state. Therefore the mathematical quantity that describes the
microphysical resonance state in a scattering experiment cannot be
given by $|z^-_R\rangle\langle^- z_R|$, but must be
given by something like $|z^\gamma_R\rangle\langle^+z_R|$.
The probability to observe $\psi^-$ in the prepared state
$\phi^+$, independently of how the effect of $\phi^+$ is carried to
the detector $\psi^-$, is given by the S-matrix element~(\ref{eq:outmatrix}),
$|(\psi^-,\phi^+)|^2$. The probability amplitude that this effect is
carried by the microphysical resonance state is then given by its pole
term $(\psi^-,\phi^+)_{\rm P.T.}$.\\
In analogy to the decay experiment one can also compare these
probabilities at different times. For this purpose 
one translates the observable $\psi^-$ in time by an amount $t\geq 0$,
\begin{equation}\label{g65}
	\psi^-\longrightarrow\psi^-(t)=e^{iHt}\psi^-;\hspace{.3in}t\geq0
\end{equation}
Physically, this would correspond to turning on the detector at a time
$t\geq0$ later than for $\psi^-$. One obtains
\begin{eqnarray}\nonumber
	\left(\psi^-(t),\phi^+\right)_{\rm P.T.}
	&=&\,-2\pi\Gamma\,\langle e^{iHt}\psi^-|z_R^\gamma\rangle 
	\langle^+z_R|\phi^+\rangle\\\label{g66}\label{WG66a}\label{ded66}
	&=&\,-2\pi\Gamma \, e^{-iz_Rt}\langle \psi^-| z_R^\gamma\rangle 
	\langle^+ z_R | \phi^+ \rangle\\\nonumber
	&=&e^{-iE_Rt}e^{-\Gamma t/2}\left(\psi^-,\phi^+\right)_{\rm P.T.}
\end{eqnarray}
This means that the time dependent probability, due to the first order
pole term, to measure the observable $\psi^-(t)$ in the state $\phi^+$
is given by the exponential law:
\begin{equation}\label{WG66b}
	|(e^{iHt}\psi^-,\phi^+)_{\rm P.T.}|^2 
	=e^{-\Gamma t} \, |(\psi^-,\phi^+)_{\rm P.T.}|^2. 
\end{equation}
This is as one would expect it if the action of the preparation
apparatus on the registration apparatus is carried by an
exponentially decaying microsystem (resonance) described by
a Gamow vector.

Therewith we have seen that there are two ways  in which a resonance
associated with a first order pole of the S-matrix ($r=1$)
can appear in experiments, and therefore there are two different forms to
represent the decaying Gamow state:
\begin{subeqnarray}
	&{\rm in~a~decay~experiment:}\hspace{16mm}
	&|z^-_R\rangle\langle^- z_R|
	\label{68b}\label{g68b}\\\label{68a}\label{g68a}
	&{\rm and~in~a~scattering~experiment:}&|z^\gamma_R\rangle\langle^+z_R|
\end{subeqnarray}
An analogous statement holds for the Gamow states 
associated with the pole in the upper half-plane.
The first representation is the one used in the S-matrix when one
calculates the cross section; the second representation is the 
one used when one calculates the Golden Rule (decay rate). 
In contrast to von Neumann's formulation where a given state
(representing an ensemble prepared by the preparation apparatus) is
always described by one and the same density operator
$|f\rangle\langle f|$, the
representation of the microphysical state in the RHS formulation depends upon
the kind of experiment one performs.
That a theory of the microsystems must include the methods of the
experiments has previously been emphasized by
G.~Ludwig. 
\\

We will now discuss the mathematical representations of 
Gamow states associated with higher
order poles of the S-matrix ($r>1$) under the two aspects above.
\\
In analogy to the case for $r=1$, we conjecture that the
representation for the microphysical system in the resonance
scattering experiment is already determined by the
pole term~(\ref{bohm41}), and is therefore given~by 
\begin{eqnarray}\label{g69a}
	W_{\rm P.T.}&=&2\pi\Gamma\sum_{n=0}^{r-1}\begin{pmatrix}r\\
	n+1\end{pmatrix}(-i)^n\frac{\Gamma^n}{n!}\sum_{k=0}^n
	\begin{pmatrix}n\\k\end{pmatrix}|z^\gamma_R\hin^{(k)}
	\;^{(n-k)}\vor^+z_R|\\
	&=&2\pi\Gamma\sum_{n=0}^{r-1}\begin{pmatrix}r\\n+1\end{pmatrix}
	(-i)^nW^{(n)}_{\rm P.T.}\nonumber
\end{eqnarray} 
up to a normalization factor which will have to be determined by
normalizing the overall probability to 1.
Here we define the operator
\begin{eqnarray}\label{g69b}
	W^{(n)}_{\rm P.T.}=\frac{\Gamma^n}{n!}\sum_{k=0}^n\begin{pmatrix}n\\
	k\end{pmatrix}|z^\gamma_R\hin^{(k)}\;^{(n-k)}\vor^+z_R|\;.
\end{eqnarray}
We hypothesize that the microphysical state from higher order poles
connected with the decay experiment has the same structure as the
microphysical state~(\ref{g69a}), which is certainly in agreement 
with the first order case~(\ref{68b}a) in comparison with~(\ref{68a}b),
\begin{eqnarray}\label{g70}
	W&=&2\pi\Gamma\sum_{n=0}^{r-1}\begin{pmatrix}
	r\\n+1\end{pmatrix}(-i)^n\frac{\Gamma^n}{n!}\sum_{k=0}^n
	\begin{pmatrix}n\\k\end{pmatrix}|z^-_R\hin^{(k)}\;^{(n-k)}
	\vor^- z_R|\\
	&=&2\pi\Gamma\sum_{n=0}^{r-1}\begin{pmatrix}r\\n+1\end{pmatrix}
	(-i)^nW^{(n)}\nonumber
\end{eqnarray}
For $r=1$, this reduces to~(\ref{68b}a).
It should be mentioned that mathematically there is an important
difference between~(\ref{g69a}) and~(\ref{g70}) because
the $\langle\psi^-|z^\gamma\hin^{(k)}\,^{(n-k)}\vor^+z|\phi^+\rangle$ 
are analytic functions for $z$ in the lower half-plane, whereas the 
$\langle\psi^-_1|z^-\hin^{(k)}\,^{(n-k)}\vor^- 
z|\psi^-_2\rangle$ are not.
\\
Whether the microphysical state of the (hypothetical) quasistationary
microphysical system is always represented by the mathematical
object~(\ref{g70}) or whether also each individual 
\begin{eqnarray}\label{operator3.12}
	\hspace{2.3cm}W^{(n)}&=&\frac{\Gamma^n}{n!}\sum_{k=0}^n
	\begin{pmatrix}n\\k\end{pmatrix}|z^-_R\hin^{(k)}\;^{(n-k)}
	\vor^- z_R|\;;\hspace{.3in}n=0,1,\dots,r-1\;,
\end{eqnarray}
has a separate physical meaning, cannot be said at this point.

This means that the conjectural physical state associated with the
$r$-th order pole is a mixed state $W$, all of whose components
$W^{(n)}$, except for the zeroth component $W^{(0)}$, cannot be
reduced further into ``pure'' states given by dyadic products like
$|z^-_R\hin^{(k)}\,^{(k)}\vor^- z_R|$. This is quite consistent
with our earlier remark that the label $k$ is not a quantum number
connected with an observable (like the suppressed labels
$b_2,\dots,b_n$). Therefore a ``pure state'' with a definite value of
$k$, like $|z^-_R\hin^{(k)}\,^{(k)}\vor^- z_R|$, $k\geq1$, does
not make sense physically. A physical interpretation could only be
given to the whole $r$-dimensional space ${\cal M}_{z_R}$, 
(\ref{bohm48}). The individual $W^{(n)}$, {$n=0,~1,~2,\dots,r-1$}, act in the
subspaces ${\cal M}^{(n)}_{z_R}\subset{\cal M}_{z_R}$ which are spanned
by Gamow vectors of order $0,~1,\dots,n\,$ (Jordan vectors of degree
$n+1$, i.e. $(H^\times-z_R)^{n+1}{\cal M}^{(n)}_{z_R}=0$). Here the
question is whether
there could be a physical meaning to each $W^{(n)}$ separately, or
whether only the particular mixture $W$ given by~(\ref{g70}) can occur
physically.

Though the quantities $|z^-_R\hin^{(k)}\,^{(k)}\vor^- z_R|$ will
have no physical meaning, even if higher order poles exist, they have
been considered in the literature
(Antoniou and Gadella, 1995; Golodberger and Watson;
Newton, 1982; Goldhaber, 1968), and their time evolution is
calculated in a straightforward way:
\begin{eqnarray}\label{purestate2}\label{g75}
	&&\hspace{-1cm}\left(e^{iHt}\right)^\times
	|z^-_R\hin^{(k)}\,^{(k)}
	\vor^- z_R|e^{iHt}=\\
	&&=e^{-\Gamma t}\sum^k_{l=0}\,\sum^k_{m=0}
	\begin{pmatrix}k\\l\end{pmatrix}
	\begin{pmatrix}k\\m\end{pmatrix}
	(-it)^l(it)^m
	|z^-_R\hin^{(k-l)}\,^{(k-m)}\vor^- z_R |\;.\nonumber
\end{eqnarray}
This state operator 
shows the additional polynomial time dependence that has always
been considered an obstacle to the use of higher order poles for
quasistationary states. A polynomial time dependence should have shown
up in many {experi}ments.\\

We now derive the time evolution of the microphysical state 
operators~(\ref{g70}) from higher order poles of the S-matrix
using the time evolution obtained for
the Gamow-Jordan vector. It will turn out that the
operator~(\ref{operator3.12}) and therewith~(\ref{g70}) 
have a purely exponential time evolution.
Inserting~(\ref{time1}) and~(\ref{time2}) into
\begin{eqnarray}\label{b75}
	W^{(n)}(t)=\left(e^{iHt}\right)^\times W^{(n)}e^{iHt}
	= \sum_{k=0}^n \, 
	      	\begin{pmatrix} 
			n \\ k 
		\end{pmatrix} \,
	\left(e^{iHt}\right)^\times |z^-_R \hin^{(k)} \,^{(n-k)} 
	\vor^- z_R |\,e^{iHt},\hspace{.2in}t\geq0\;,
\end{eqnarray}
we get
\begin{equation}\nonumber
	W^{(n)}(t)=e^{-iz_Rt}e^{iz^*_Rt}
	\frac{\Gamma^n}{n!}\sum_{k=0}^n
	\sum_{l=0}^k\sum_{m=0}^{n-k} 
	\begin{pmatrix} n \\ k \end{pmatrix}
	\begin{pmatrix} k \\ l \end{pmatrix}
	\begin{pmatrix} n\! -\! k \\ m \end{pmatrix}
	(-it)^{k-l}(it)^{n-k-m}
	|z^-_R\hin^{(l)}\,^{(m)}\vor^- z_R|\;.	
\end{equation}
After reordering the summations and the terms in the binomial
coefficients one can separate out the dyades,
\begin{equation}\nonumber
	W^{(n)}(t)=e^{-\Gamma t}\frac{\Gamma^n}{n!}\sum_{m=0}^n
	\begin{pmatrix} n \\ m \end{pmatrix}
	\sum_{l=0}^{n-m}\begin{pmatrix} \!n\!-m\!\\ l \end{pmatrix}
	|z^-_R\hin^{(l)}\,^{(m)}\vor^- z_R|\sum_{k=l}^{n-m}
	\begin{pmatrix} \!n\!-\!m\!-\!l\\ k\!-\!l \end{pmatrix}
	(-it)^{k-l}(it)^{n-k-m}\;.
\end{equation}
Since the indices labeling the Gamow-Jordan vectors do not depend
upon $k$, the sum over $k$ may be performed using the binomial formula
\begin{equation}\nonumber
	\sum_{k=l}^{n-m}\, 
	\begin{pmatrix}\!n\!-\!m\!-\!l\!\\k\!-\!l\end{pmatrix}	
	(-it)^{k-l}\, (it)^{n-k-m}
	= (it-it)^{n-m-l} = 
	\left\{
		\begin{array}{c} 
			1 \quad \mbox{for $l=n-m$} \\ 
			0 \quad \mbox{for $l\neq n-m$} 
		\end{array} 
	\right\}=\delta_{l,n-m}
\end{equation}
and one gets 
\begin{equation}\label{b82'}
	W^{(n)}(t)=e^{-\Gamma t}\frac{\Gamma^n}{n!}\,\sum_{m=0}^n 
      	\begin{pmatrix}n\\m\end{pmatrix}  
  	|z^-_R \hin^{(n-m)} \,^{(m)}\vor^- z_R |
	=e^{-\Gamma t}\,W^{(n)}(0)\,;\hspace{.2in}t\geq0\;.
\end{equation}
This means that the non-reducible (i.e. ``mixed'')
state operator $W^{(n)}$ of~(\ref{operator3.12}) 
has a simple exponential semigroup time
evolution, and also the operator $W$ of~(\ref{g70}) being a linear
combination of the $W^{(n)}$, 
\begin{equation}\label{b83}
	W(t)\equiv \left(e^{iHt}\right)^\times 
	We^{iHt}=e^{-\Gamma t}W\;,\quad t\geq0\,.
\end{equation}
It turns out that the
operator~(\ref{operator3.12}) is the only operator in ${\cal
M}_{z_R}^{(n)}$ formed by the dyadic products $|z^-_R \hin^{(m)}
\,^{(l)} \vor^- z_R |$ with $~m,~l=0,~1,\cdots,n$, which has a purely
exponential time evolution, thus being distinguished from all other
operators in ${\cal M}_{z_R}^{(n)}$. 
Thus we have seen that the state operator which we conjecture from the
$r$-th order pole term describes a non-reducible ``mixed''
microphysical decaying state which obeys an exact exponential decay law.

In analogy to~(\ref{g66}) one can also calculate the time evolution of
the operators~(\ref{g69a}) and~(\ref{g69b}), but their time evolution
will always have an additional polynomial time dependence besides the
exponential. 

\section{General Form of the Exponentially Decaying Gamow State}
\label{sec:general}
\setcounter{equation}{0}

In this section we want to discuss the converse of the above
reasoning, where we conjectured the density operator for higher order
decaying Gamow states and derived a purely exponential time evolution.
We ask the question: If we require exponentially decaying time evolution
for a Gamow state operator formed by dyadic products of vectors in
${\cal M}_{z_R}$, what is the most general form of such an operator?

We denote the most general form of $W$ of finite dimension $j\in I\!\!N$ by
\begin{equation}\label{chris1}
	W^\square_{(j)}(0)=\sum_{k=0}^j\sum_{h=0}^j
	A_{hk}|z^-_R\hin^{(k)}\,^{(h)}\vor^- z_R|\;.
\end{equation}
with arbitrary coefficients $A_{hk}$. 
We want to change the order of summation from the 
states of the form $|z^-_R\hin^{(k)}\,^{(h)}\vor^- z_R|$ 
to the states of the form $|z^-_R\hin^{(k)}\,^{(n-k)}\vor^- z_R|$. 
Changing the label $h=n-k$ we write this sum as
\begin{equation}\label{chripe}
	W^\square_{(j)}(0)=\sum_{k=0}^j\sum_{n=k}^{j+k}
	A_{n-k,k}|z^-_R\hin^{(k)}\;^{(n-k)}\vor^- z_R|\;.
\end{equation}
Switching the order of $k$ and $n$ divides~(\ref{chripe}) into two sums,
\begin{equation}
	\sum_{k=0}^j\sum_{n=k}^{j+k}=\sum_{n=0}^j\sum_{k=0}^n
	+\sum_{n=j+1}^{2j}\sum_{k=n-j}^j\;.
\end{equation}
We define $W^\square_{(j)}=W_{(j)}^\Bigtriangleright 
+W_{(j)}^\Bigtriangleleft$, such that
\begin{eqnarray}\label{chris2}
	W_{(j)}^\Bigtriangleright&=&\sum_{n=0}^j\sum_{k=0}^n
	A_{n-k,k}|z^-_R\hin^{(k)}\;^{(n-k)}\vor^- z_R|\;,\\
	W_{(j)}^\Bigtriangleleft&=&\sum_{n=j+1}^{2j}\sum_{k=n-j}^j
	\label{chris3}
	A_{n-k,k}|z^-_R\hin^{(k)}\;^{(n-k)}\vor^- z_R|\;.
\end{eqnarray}
In the following, we calculate the coefficients of 
$W_{(j)}^\Bigtriangleright$ and give an argument why this suffices to
conlude that the coefficients of 
$W_{(j)}^\Bigtriangleleft$ all turn out to be zero.

The time dependence of
$W^\Bigtriangleright$ is given, using~(\ref{time1}) and~(\ref{time2}), by 
\begin{eqnarray}\nonumber
  	W_{(j)}^\Bigtriangleright(t)
	&=&\left(e^{iHt}\right)^{\times}W_{(j)}^\Bigtriangleright e^{iHt}=
  	\sum_{n=0}^j\sum_{k=0}^nA_{n-k,k}\left(e^{iHt}\right)^{\times}
  	|z_R^-\hin^{(k)}\,^{(n-k)}\vor^- z_R|e^{iHt}\\
  	&=&e^{-\Gamma t}\sum_{n=0}^j\sum_{k=0}^n\sum_{l=0}^k\sum_{m=0}^{n-k}
  	A_{n-k,k}\begin{pmatrix}k\\l\end{pmatrix}\begin{pmatrix}n\!-\!k\\ 
  	m\end{pmatrix}\left(-it\right)^{k-l}\left(it\right)^{n-k-m}
  	|z_R^-\hin^{(l)}\,^{(m)}\vor^- z_R|\,.\nonumber
\end{eqnarray}
Changing the order of the summations,
\begin{equation}\nonumber
  \sum_{n=0}^j
  \sum_{k=0}^n
  \sum_{l=0}^k
  \sum_{m=0}^{n-k}
  \!=\!
  \sum_{n=0}^j
  \sum_{l=0}^n
  \sum_{k=l}^n
  \sum_{m=0}^{n-k}
  \!=\!
  \sum_{l=0}^j
  \sum_{n=l}^j
  \sum_{k=l}^n
  \sum_{m=0}^{n-k}
  \!=\!
  \sum_{l=0}^j
  \sum_{n=l}^j
  \sum_{m=0}^{n-l}
  \sum_{k=l}^{n-m}
  \!=\!
  \sum_{l=0}^j
  \sum_{m=0}^{j-l}
  \sum_{n=l+m}^j
  \sum_{k=l}^{n-m},
\end{equation}
allows the dyadic products, which are linearly independent operators, to be
factored out of the sums over terms in which they appear as common factors:
\begin{eqnarray}\nonumber
  W_{(j)}^\Bigtriangleright(t)&=&
  e^{-\Gamma t}
  \sum_{l=0}^j
  \sum_{m=0}^{j-l}
  \sum_{n=l+m}^j
  \sum_{k=l}^{n-m}
  A_{n-k,k}
  \begin{pmatrix}k\\l\end{pmatrix}\begin{pmatrix}n\!-\!k\\m\end{pmatrix}
  \left(-it\right)^{k-l}
  \left(it\right)^{n-k-m}
  |z_R^-\hin^{(l)}
  \,^{(m)}\vor^- z_R|\\\nonumber
  &=&
  e^{-\Gamma t}
  \sum_{l=0}^j
  \sum_{m=0}^{j-l}
  |z_R^-\hin^{(l)}
  \,^{(m)}\vor^- z_R|
  \sum_{n=l+m}^j
  \sum_{k=l}^{n-m}
  A_{n-k,k}
  \begin{pmatrix}k\\l\end{pmatrix}\begin{pmatrix}n\!-\!k\\m\end{pmatrix}
  \left(-it\right)^{k-l}
  \left(it\right)^{n-k-m}\\\nonumber
  &=&
  e^{-\Gamma t}
  \sum_{l=0}^j
  \sum_{m=0}^{j-l}
  |z_R^-\hin^{(l)}
  \,^{(m)}\vor^- z_R|
  \sum_{n=l+m}^j
  \left(it\right)^{n-m-l}
  \sum_{k=l}^{n-m}
  A_{n-k,k}
  \begin{pmatrix}k\\l\end{pmatrix}\begin{pmatrix}n\!-\!k\\m\end{pmatrix}
  \left(-1\right)^{k-l}\,.
\end{eqnarray}

The operator 
$W^\Bigtriangleright(t)$ will decay according to the pure exponential
$e^{-\Gamma t}$ if and only if all terms involving additional powers of $t$
cancel.  All terms involving additional powers of $t$ will cancel if and only
if the coefficients $A_{n-k,k}$ satisfy the conditions
\begin{equation}\label{conditions}
  0
  =
  \sum_{k=l}^{n-m}
  A_{n-k,k}
  \begin{pmatrix}k\\l\end{pmatrix}\begin{pmatrix}n\!-\!k\\m\end{pmatrix}
  \left(-1\right)^{k-l}
  \hskip 5mm\hbox{for}\hskip 5mm
  \left\{\begin{array}{l}
  l\in\{0,\cdots,j-1\}, \\
  m\in\{0,\cdots,j-1-l\}, \\
  n\in\{m+l+1,\cdots,j\}, \end{array}
  \right.
\end{equation}
The simplest of these conditions are those for which $n=m+l+1$, i.e., those for
which $m=n-l-1$, because they are the only conditions that involve sums over
only two values of $k$:
\begin{eqnarray}\nonumber
  0
  &=&
  \sum_{k=l}^{l+1}
  A_{n-k,k}
  \begin{pmatrix}k\\l\end{pmatrix}
  \begin{pmatrix}n\!-\!k\\n\!-\!l\!-\!1\end{pmatrix}
  \left(-1\right)^{k-l}
  \\\nonumber
  &=&
  A_{n-l,l}
  \begin{pmatrix}n\!-\!l\\ n\!-\!l\!-\!1\end{pmatrix}
  -
  A_{n-l-1,l+1}
  \begin{pmatrix}l\!+\!1\\l\end{pmatrix}
  \hskip 10mm\hbox{for}\hskip 10mm
  \left\{\begin{array}{l}
  n\in\{1,\cdots,j\}, \\
  l\in\{0,\cdots,n-1\}\end{array}
  \right.
\end{eqnarray}
or, equivalently,
\begin{equation}\nonumber
  A_{n-k+1,k-1}
  \begin{pmatrix}n\!-\!k\!+\!1\\n\!-\!k\end{pmatrix}
  =
  A_{n-k,k}
  \begin{pmatrix}k\\k\!-\!1\end{pmatrix}
  \hskip 12mm\hbox{for}\hskip 10mm
  \left\{\begin{array}{l}
  n\in\{1,\cdots,j\}, \\
  k\in\{1,\cdots,n\} \end{array}
  \right.
\end{equation}
or, equivalently,
\begin{equation}\nonumber
  A_{n-k,k}
  =
  \frac{(n\!-\!k\!+\!1)!(k\!-\!1)!}{(n\!-\!k)!k!}
  A_{n-k+1,k-1}
  \hskip 15mm\hbox{for}\hskip 10mm
  \left\{\begin{array}{l}
  n\in\{1,\cdots,j\}, \\
  k\in\{1,\cdots,n\}. \end{array}
  \right.
\end{equation}
These conditions relate pairs of coefficients $A_{n-k,k}$ having the
same values of $n$ and successive values of $k$.  For fixed
$n\in\{1,2,\cdots,j\}$, they may be used recursively to show that
$A_{n-k,k}$ must equal $A_{n,0}$ multiplied by 
the binomial coefficient ${n\choose k} \equiv \frac{n!}{k!(n-k)!}$:
\begin{eqnarray}\nonumber
  A_{n-k,k}
  &=&
  \left[\frac{(n\!-\!k\!+\!1)!(k\!-\!1)!}{(n\!-\!k)!k!}\right]
  \left[\frac{(n\!-\!k\!+\!2)!(k\!-\!2)!}{(n\!-\!k\!+\!1)!(k\!-\!1)!}\right]
  \cdots
  \left[\frac{(n\!-\!1)!1!}{(n\!-\!2)!2!}\right]
  \left[\frac{n!0!}{(n\!-\!1)!1!}\right]
  A_{n,0}
  \\
  &=&
  \frac{n!0!}{(n\!-\!k)!k!}
  A_{n,0}
  =
  {n\choose k}
  A_{n,0}
  \hskip 10mm\hbox{for}\hskip 10mm
  \left\{\begin{array}{l}
  n\in\{1,\cdots,j\}, \\
  k\in\{1,\cdots,n\}. \end{array}
  \right.				\label{binomial}
\end{eqnarray}
Substituting this result into the full set of
conditions~(\ref{conditions}), using the identity
\begin{equation}\nonumber
  {n\choose k}
  {k\choose l}
  {n\!-\!k\choose m}
  =
  {n\choose m}
  {n\!-\!m\choose l}
  {n\!-\!m\!-\!l\choose k\!-\!l},
\end{equation}
and then using the binomial formula gives
\begin{eqnarray}\nonumber
  0
  &=&
  A_{n,0}
  \sum_{k=l}^{n-m}
  {n\choose k}
  {k\choose l}
  {n\!-\!k\choose m}
  \left(-1\right)^{k-l}
  \\\nonumber
  &=&
  A_{n,0}
  {n\choose m}
  {n\!-\!m\choose l}
  \sum_{k=l}^{n-m}
  {n\!-\!m\!-\!l\choose k\!-\!l}
  \left(-1\right)^{k-l}
  \\\nonumber
  &=&
  A_{n,0}
  {n\choose m}
  {n\!-\!m\choose l}
  \sum_{k-l=0}^{n-m-l}
  {n\!-\!m\!-\!l\choose k\!-\!l}
  1^{n-m-k}
  \left(-1\right)^{k-l}
  \\\nonumber
  &=&
  A_{n,0}
  {n\choose m}
  {n\!-\!m\choose l}
  (1-1)^n
  \\\nonumber
  &=&
  A_{n,0}
  {n\choose m}
  {n\!-\!m\choose l}
  0^n
  \hskip 10mm\hbox{for}\hskip 10mm
  \left\{\begin{array}{l}
  l\in\{0,\cdots,j-1\}, \\
  m\in\{0,\cdots,j-1-l\}, \\
  n\in\{m+l+1,\cdots,j\},\end{array}
  \right.
\end{eqnarray}
which shows that the remaining conditions are automatically satisfied
by~(\ref{binomial}) 
without placing any further conditions on the coefficients $A_{n,0}$.  The
coefficients $A_{n,0}$, for $n\in\{1,\cdots,j\}$, and also the coefficient
$A_{0,0}$, remain completely arbitrary.

We conclude that a linear combination of dyadic products 
$|z_R^-\hin^{(l)}\,^{(m)}\vor^- z_R|$ 
decays according to the pure exponential
$e^{-\Gamma t}$ if and only if it is of the form
\begin{equation}\label{con2}
  \sum_{n=0}^j
  A_{n,0}
  \sum_{k=0}^n
  {n\choose k}
  |z_R^-\hin^{(k)}
  \,^{(n-k)}\vor^- z_R|
\end{equation}
with arbitrary coefficients $A_{n,0}$.

Now, coming back to the argument at the beginning of this section:
We have shown that for arbitrary $j$, the operator $W_{(j)}^\Bigtriangleright$
depends only on the choice of the coefficients $A_{n,0}$. Since $j$
was chosen arbitrarily, we can also
take an operator $W^\Bigtriangleright$ of dimension $2j$, such that it
contains at least all the terms belonging to the operator 
$W_{(j)}^\square$ of~(\ref{chris1}) and some additional terms which we
set zero by choice of the coefficients $A_{hk}=0$ for $h,k>j$,
\begin{eqnarray}
	W_{(j)}^\square&=&W_{(2j)}^\Bigtriangleright
	\hspace{1cm}{\rm with}~A_{hk}=0~{\rm for}~k>j~{\rm or}~h>j\\
	&=&\sum_{k=0}^{2j}\sum_{h=0}^{2j-k}\nonumber
	A_{hk}|z^-_R\hin^{(k)}\;^{(h)}\vor^-
	z_R|\;;\hspace{1cm}{\rm with}~A_{hk}=0~{\rm for}~k>j~{\rm or}~h>j\;.
\end{eqnarray}
If for $n>j$ the $A_{n,0}=0$ then we know that, according to~(\ref{binomial}),
all the other terms $A_{n-k,k}$ are zero. Since they make up the 
the operator $W_{(j)}^\Bigtriangleleft$ of~(\ref{chris3}), we conclude
that the coefficients of $W_{(j)}^\Bigtriangleleft$ are zero, as
demanded above.

Comparing ~(\ref{con2}) with arbitrary coefficients $A_{n,0}$ 
to the Gamow state operator~(\ref{g70}) suggested by the pole term, 
one sees that the structure is the same with $j=r-1$ and 
\begin{equation}
	A_{n,0}=2\pi\Gamma\begin{pmatrix}
	r\\n+1\end{pmatrix}(-i)^n\frac{\Gamma^n}{n!}\;.
\end{equation}

\section{Conclusion}
\label{sec:conlusion}
\setcounter{equation}{0}

Gamow vectors that can describe resonances and decaying states from
first order poles of the S-matrix have been known for two decades. In
this paper, we discussed their generalization to Gamow vectors
describing resonances from higher order S-matrix poles. This led to a
set of $r$ higher order Gamow vectors associated to a pole of
multiplicity $r$, which are Jordan vectors to a self-adjoint
Hamiltonian with complex eigenvalue $E_R-i\Gamma/2$. They are basis
elements of a generalized eigenvector expansion which suggests the
form of a microphysical state associated with this higher order
resonance pole. This microphysical state is a mixture of non-reducible
components, and in spite of the fact that the higher order Gamow
vectors have an additional polynomial time dependence, this
microphysical state obeys a purely exponential decay law. We showed
that this state operator has the same structure as the 
most general from constructed from decaying higher order Gamow vectors
that leads to an exponential decay law.\\
Katznelson, 1980; Bhamathi and Sudarshan, 1995; Br\"{a}ndas and
Dreismann, 1987; Antoniou and Tasaki, 1993).
It has been shown that Jordan blocks arise naturally from higher order
S-matrix poles and represent a self-adjoint Hamiltonian
by a complex matrix in a finite
dimensional subspace contained in the rigged Hilbert space. Although
higher order S-matrix poles are not excluded theoretically, there has
been so far very little experimental evidence for their existence,
because they were always believed to have polynomial time dependence.
Our results suggest that the
empirical objection to the existence of higher order poles of the
S-matrix does not rule out the possibility of exponentially decaying
states constructed from higher order Gamow vectors.

\section*{Acknowledgment}
\label{acknowledgment}
\setcounter{equation}{0}

I am very happy to have been invited to Peyresq, and I wish to express
my sincere gratitude to Mario Castagnino
and Edgard and Diane Gunzig, who organized the meeting.
A special thank you goes to Mady Smets
for her extraordinary hospitality and that of her fine staff.

A little anecdote goes with this paper: The 
fourth section of this paper evolved as a reply to a question 
from one of the members of the audience during 
my presentation. I would like to express my gratitude to this person 
in cognito.

Finally, I would like to acknowledge the members of my research
group, A.~Bohm, M.~Gadella, G.~Pronko, M.~Loewe, P.~Patuleanu, and
S.~Wickramasekara, whose gratious help made this work possible and enjoyable.

\section*{References}

\footnotesize{
\begin{enumerate}

\item[]
Antoniou,~I., and Gadella,~M. (1995). International Solvay Institute,
  Brussels, Preprint. [Results of this preprint were published in:
  A.~Bohm {\em et al.} (1995). 
  {\em Reports on Mathematical Physics}~{\bf 36}, 245.]

\item[]
Antoniou,~I., and Tasaki,~S. (1993). {\em International Journal of
Quantum Chemistry}~{\bf 46}, 425.

\item[]
Baumg{\"{a}}rtel,~H. (1984). {\em Analytic Perturbation Theory for Matrices and
  Operators}, Chap.~2, Birkh\"{a}user, Basel.

\item[]
Bhamathi, G., and Sudarshan, E.~C.~G. (1996). International Journal of
Modern Physics. {\bf B 10}, 1531.

\item[]
Bohm,~A. (1979). {\em Letters of Mathematical Physics}~{\bf 3}, 455. 

\item[]
Bohm,~A. (1980). {\em Journal of Mathematical Physics}~{\bf 21}(5), 1040.

\item[]
Bohm,~A. (1981). {\em Journal of Mathematical Physics}~{\bf 22}(12), 2813.

\item[]
Bohm,~A., and Gadella,~M. (1989). 
  {\em Dirac Kets, Gamow Vectors and Gel'fand Triplets},
  \newblock Springer, Berlin.

\item[]
Bohm,~A. (1993). {\em Quantum~Mechanics}, {$3^{rd}$}~ed.,
  Springer,~Berlin.

\item[]
Bohm,~A., Maxson,~S., Loewe,~M., and Gadella,~M. (1997). To appear in
  Physica A. 

\item[]
Br{\"{a}}ndas,~E.~J., and Chatzidimitriou-Dreismann,~C.~A. (1987).
  {\em Resonances},
  {\em Lecture Notes in Physics} {\bf 325}, 480, 
  E.~J. Br{\"{a}}ndas and N.~Elander, eds.,
  Springer, Berlin.

\item[]
Dothan,~Y., and Horn,~D. (1970). {\em Physical Review D}, {\bf 1}, 6.

\item[]
Duren,~P.~L. (1970). {\em Theory of ${\cal H}^p$ Spaces},
  \newblock Academic, New~York.

\item[]
Gantmacher,~F.~R. (1959). {\em Theory of Matrices}, section~VII.7,
  Chelsea, New York. 

\item[]
Gelfand,~I.~M., and Vilenkin,~N.~Ya. (1964).
  {\em Generalized Functions}, vol.~IV,
  Academic, New York.

\item[]
Goldberger,~M.~L., and Watson,~K.~M. (1964).
  {\em Collision~Theory}, Wiley, New~York.

\item[]
Goldberger,~M.~L., and Watson,~K.~M. (1964). {\em Physical Review} 
{\bf 136} B1472.

\item[]
Goldhaber,~A.~S. (1968). {\em Meson~Spectroscopy}, p.~297, 
  C.~Baltay and A.~H. Rosenfeld, eds.,
  \newblock Benjamin, New York.

\item[]
Hoffman, K. (1962). {\em Banach Spaces of Analytic Functions}, 
  Prentice-Hall, Englewood Cliffs, N.~J. 

\item[]
Kato,~T. (1966). {\em Perturbation Theory for Linear Operators}, 
  Springer, Berlin.

\item[]
Katznelson,~E. (1980). {\em Journal of Mathematical Physics}~{\bf 21}, 1393.

\item[]
Lancaster,~P., and Tismenetsky,~M. (1985).
  {\em Theory of Matrices}, $2^{nd}$ ed.,
  Academic, New York.
(See also ref. [20] in ref. [10]).

\item[]
Ludwig,~G. {\em Foundations of Quantum Mechanics}, 
  \newblock Vol.~I, Springer, Berlin (1983);
  \newblock Vol.~II (1985).

\item[]
Ludwig,~G. {\em An Axiomatic Basis for Quantum Mechanics},
  \newblock Vol.~I, Springer, Berlin (1985); Vol.~II (1987).

\item[] 
Lukierski,~J. (1967). {\em Bul.~Acad. Polish Science}~{\bf 15}, 223.

\item[]
Mondrag\'{o}n,~A. (1994). {\em Physics~Letters~B} {\bf 326}, 1 and
references thereof.

\item[]
Newton,~R.~G. (1982).
  {\em Scattering Theory of Waves and Particles}, 2$^{\rm nd}$ ed.,
  Springer, New York.

\item[]
Stodolsky,~L. (1970). {\em Experimental Meson Spectroscopy}, p.~395,
  C.~Baltay and A.~H. Rosenfeld, eds., Columbia, New York.

\end{enumerate}}

\bibliographystyle{ieeetr}

\end{document}